\documentclass[10pt,letterpaper]{article}
\usepackage{opex3}
\usepackage{amsmath}
\usepackage{bm}

\begin{document}

\title{Local orbital-angular-momentum dependent surface states with topological protection}

\author{Menglin L. N. Chen$^1$, Li Jun Jiang$^{1,4}$, Zhihao Lan$^2$, and Wei E. I. Sha$^{3,5}$}

\address{$^1$Department of Electrical and Electronic Engineering, The University of Hong Kong, Hong Kong\\ 
$^2$Department of Electronic and Electrical Engineering, University College London, United Kingdom\\
$^3$Key Laboratory of Micro-nano Electronic Devices and Smart Systems of Zhejiang Province, College of Information Science and Electronic Engineering, Zhejiang University, Hangzhou 310027, China}

\email{$^4$jianglj@hku.hk} 
\email{$^5$weisha@zju.edu.cn}


\begin{abstract}
Chiral surface states along the zigzag edge of a valley photonic crystal in the honeycomb lattice are demonstrated. By decomposing the local fields into orbital angular momentum (OAM) modes, we find that the chiral surface states present OAM-dependent unidirectional propagation characteristics. Particularly, the propagation directivities of the surface states are quantified by the local OAM decomposition and are found to depend on the chiralities of both the source and surface states. These findings allow for the engineering control of the unidirectional propagation of electromagnetic energy without requiring an ancillary cladding layer. Furthermore, we examine the propagation of the chiral surface states against sharp bends. It turns out that although only certain states successfully pass through the bend, the unidirectional propagation is well maintained due to the topology of the structure.
\end{abstract}



\bibliographystyle{osajnl}
\bibliography{reference}


\section{Introduction}

As a new phase of matter, topological insulators that preserve time-reversal symmetry have been extensively studied \cite{kane2010colloquium, sczhang2011TI}. In a two-dimensional (2D) topological insulator, there exist two edge channels associated with different spin orientations of the occupying electrons~\cite{kane2005quantum, sczhang2007quantum}, forming quantum spin Hall (QSH) states. Interestingly, topological insulators have their analogues in electromagnetics~\cite{ozawa2019topological}. To mimic the spin degree of freedom of electrons, different schemes in electromagnetics have been proposed and verified~\cite{hafezi2011robust, liang2013optical, Khanikaev2013photonic, Ma2015guiding, huxiao2015scheme, Khanikaev2016robust, hangzh2018visualization,liu2019topological,menglinTAP}. For example, in a simple scenario using dielectric photonic crystals (PCs) with $C_{6v}$ symmetry~\cite{huxiao2015scheme}, two degenerate modes with pseudospin-up and -down polarizations are constructed from hexagonal clusters and pseudospin-momentum locked edge states are supported at the interface between topologically nontrivial and trivial PCs. Apart from spin, valley is another degree of freedom, which provides the valley contrasting transport~\cite{dixiao2007valley}, i.e. the quantum valley Hall (QVH) effect. Correspondingly, the electromagnetic (EM) version of the valley degree of freedom has also been studied~\cite{Ma2016all, dongjw2017valley, zhangbl2017valley, zhangbl2018valley, dongjw2018tunable, wu2017direct, hangzh2018topological, zhangbl2018topologically}.

The helical nature of the edge states in the topological PCs emulating spin degree of freedom allows the unidirectional excitation and propagation of EM waves against moderate disorders and sharp bends~\cite{huxiao2015scheme, hangzh2018visualization}, which is quite different from the conventional PC waveguiding modes~\cite{jamois2003silicon}. As for the valley PCs, when there is no inter-valley scattering, the unidirectional propagation can be preserved, such as the reflection-free out-coupling into vacuum at the zigzag termination~\cite{Ma2016all} and the broadband robust transmission in the presence of sharp corners~\cite{dongjw2017valley}.

In this work, we present chiral edge states that are supported at the zigzag edge of a single valley PC in contact with air, which is very different from previous studies where edge states are supported at the interface separating two valley PCs symmetrical to each other under inversion~\cite{Ma2016all, dongjw2017valley, zhangbl2017valley, zhangbl2018valley, dongjw2018tunable, wu2017direct,  hangzh2018topological, zhangbl2018topologically}. The edge states we considered are well localized at the edge sites and evanescent both inside and outside the valley PC, and hence become surface states. We note that while surface states in PCs have been studied~\cite{joannopoulos2011PCs, Rahachou2006waveguiding, noda2009manipulation}, their possible topological protection has not been studied to the best of our knowledge. The interesting thing with these chiral surface states is that the forward and backward propagating waves can be selectively excited by choosing the chirality of the source. Furthermore, due to the topology of the valley PC, the unidirectional propagation of the surface states is maintained when there are sharp bends.

\section{Valley photonic crystals}

\begin{figure}[htbp]
	\centering
	\includegraphics[width=0.7\columnwidth]{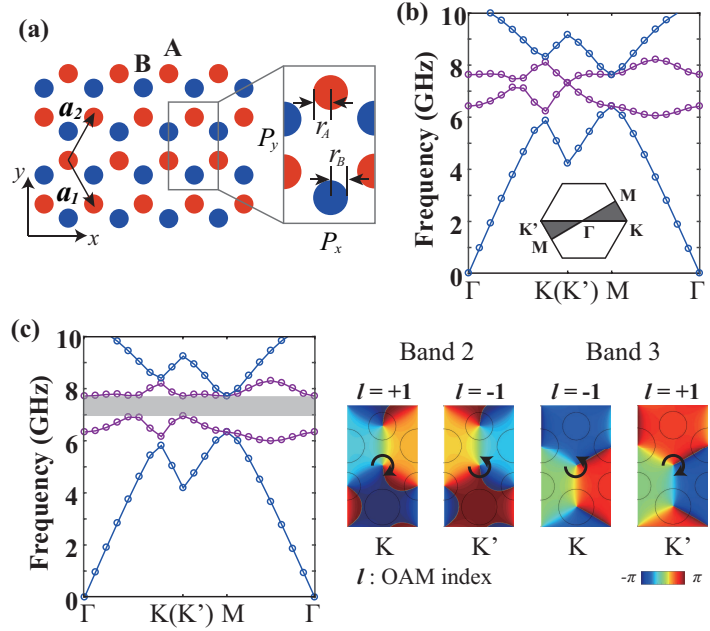}
	\caption{2D valley PC. (a) Geometry of the PC with translation vectors, $\bm{a_1}$ and $\bm{a_2}$. The inset shows the supercell holding the periodicity along the $x$ and $y$ directions. Its band structure when (b) $r_A=r_B=2.3$~mm and (c) $r_A=2$~mm, $r_B=2.6$~mm with the right panel showing the phase of $E_z$ at the $K$ and $K'$ points. The dielectric constant of the cylinders $\epsilon=12$ and $a_0=6\sqrt3$~mm.}
	\label{cyl}
\end{figure}

A 2D PC arranged in a honeycomb lattice is shown in Fig.~\ref{cyl}(a). $\bm{a_1}$ and $\bm{a_2}$ are the two translation vectors with the length of $a_0$, i.e. the lattice constant. There are two types of cylinders with the same dielectric constant $\epsilon$ but different radii $r_A$ and $r_B$. Only the transverse-magnetic (TM) modes are considered, i.e. electric field only has the out-of-plane component and magnetic field is confined to the $xoy$ plane. We build a rectangular supercell with the periodic boundary conditions imposed at the $x$ and $y$ directions. The lengths of the reassigned translation vectors are $P_x=a_0$ along $x$ and $P_y=\sqrt{3}a_0$ along $y$. Then, we calculate its band structure by using the finite-difference (FD) method~\cite{menglinPRA}.

A Dirac point is where two linear dispersion curves intersect and the corresponding crystal momenta in 2D k-space form the Dirac cone. Since Dirac cones appear at the $K$ and $K'$ points in honeycomb lattices, while the rectangular supercell is used in the band structure calculation for the convenience of FD method, the first Brillouin zone is calculated according to the two translation vectors of $\bm{a_1}$ and $\bm{a_2}$. Figure~\ref{cyl}(b) depicts the first four bands when $r_A=r_B$. There is no band gap and two Dirac cones with a two-fold degeneracy appear as expected at the $K$ and $K'$ points. While when $r_A\neq r_B$, the inversion symmetry is broken, so the Dirac cones become two valleys. The band gap is opened up from $6.95$~GHz to $7.71$~GHz in Fig.~\ref{cyl}(c). The two valleys have the same eigenfrequency but they are inequivalent. The right panel of Fig.~\ref{cyl}(c) shows the phase of the four states ($E_z$) at the $K$ and $K'$ valleys. Obviously, the electric fields within the supercells are OAM dependent. The electric fields carry an OAM of order $1$ or $-1$, whose sign is determined by the phase winding direction. The phase winding directions are opposite for the second and third bands at the same valley. Meanwhile, they are opposite at the same band but different valleys.

\section{Surface states with chirality}

Valley-dependent edge states have been observed at the interface of two valley PCs with interchanged $r_A$ and $r_B$ ~\cite{dongjw2017valley,zhangbl2018valley} and one valley PC but with an external cladding~\cite{dongjw2018tunable}. In the following, we will demonstrate the cladding-free guidance of valley-dependent edge states.

\begin{figure}[htbp]
	\centering
	\includegraphics[width=0.7\columnwidth]{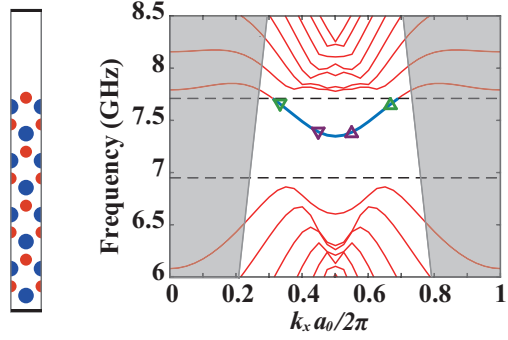}
	\caption{Geometry and band structure of the supercell composed of the valley PC in Fig.~\ref{cyl}(c). The dashed black lines indicate the band gap edges. The gray regions are where the states are extended in the air. The surface states are on the blue line.}
	\label{BG_half}
\end{figure}

The supercell of the proposed structure is drawn in Fig.~\ref{BG_half}. The parameters of the valley PC are the same as those of Fig.~\ref{cyl}(c). We adopt the same FD solver described in the previous section to solve this problem. Bloch boundary conditions are imposed on the left and right borders while two perfectly matched layers are specified at the upper and lower borders of the supercell based on the complex coordinate stretching approach~\cite{chew1994}. The band structure is calculated by sweeping $k_x$ from $0$ to $2\pi/a_0$. The black dashed lines mark the band gap calculated in Fig.~\ref{cyl}(c). The gray regions are where the states are extended in the air. The surface states should locate below the light lines and also within the band gap. Therefore, we can identify the surface states on the blue line (from $7.35$~GHz to $7.71$~GHz), on which we marked four typical states. Note that the periodicity of the structure along $x$ is much smaller than the wavelength, so only zero-order mode exists and the surface states have no loss along the $x$ direction. The green triangles are where the $K$ and $K'$ points locate and the purple triangles are two positions near $k_x a_0/2\pi=0.5$. According to the Bloch's theorem, the branch with $k_x a_0/2\pi<0.5$ corresponds to $k_x>0$, while the branch with $k_x a_0/2\pi>0.5$ corresponds to $k_x<0$. Hence, the modes represented by triangles have negative wave number, while the modes represented by inverted triangles have positive wave number.

\begin{figure}[htbp]
	\centering
	\includegraphics[width=0.7\columnwidth]{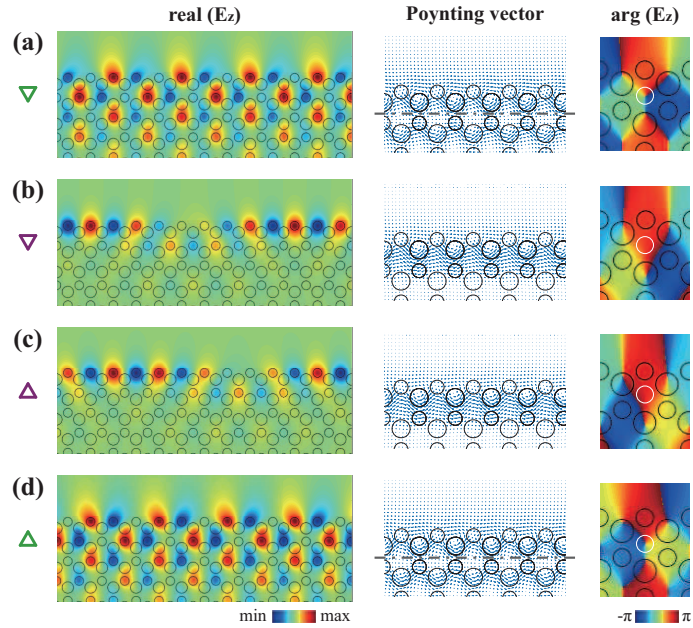}
	\caption{The surface states at the valley PC and air interface: the real-space distributions of the electric fields, time-averaged Poynting vectors, and the phase distributions of the electric fields at the four marked points on the band structure in Fig.~\ref{BG_half}. The white line indicates a full circular path around the center of the hexagon formed by the six surrounding cylinders.}
	\label{BG_half_E}
\end{figure}

The real-space distributions of the electric fields ($E_z$) and the time-averaged Poynting vectors at the four marked locations are plotted in Fig.~\ref{BG_half_E}. For the two states near $k_x a_0/2\pi=0.5$, i.e. in Figs.~\ref{BG_half_E}(b) and (c), the EM fields are pinned to the surface plane. In Figs.~\ref{BG_half_E}(a) and (d), the EM fields at the $K$ and $K'$ points extend several lattices to the bulk, because the two states are quite close to the upper boundary of band gap. From the distributions of Poynting vectors, we can see for all the four surface states, the EM energy flows from one supercell to its adjacent supercell, forming an effective wave-guiding channel. As expected, for the states on the left branch (Figs.~\ref{BG_half_E}(a) and (b)), the energy flows leftward with positive wave number and the states on the right branch (Figs.~\ref{BG_half_E}(c) and (d)) propagate rightward with negative wave number. It is worth noting that the left- and right-flowing energy paths are accompanied by half-cycle orbits, indicating an inherent chirality of the surface states. The phase distributions of $E_z$ within one hexagonal cluster are drawn on the right panel. At the $K$ and $K'$ points, the phase distributions are similar to those of the bulk states in Fig.~\ref{cyl}(c): there is a gradual phase increment around the center. Moreover, the directions of the increment are opposite for $K$ and $K'$ points. The chirality of the surface states in Figs.~\ref{BG_half_E}(b) and (c) is hard to tell because besides the OAM of order $\pm 1$, there are field components carrying no OAM.

\begin{figure}[htbp]
	\centering
	\includegraphics[width=0.7\columnwidth]{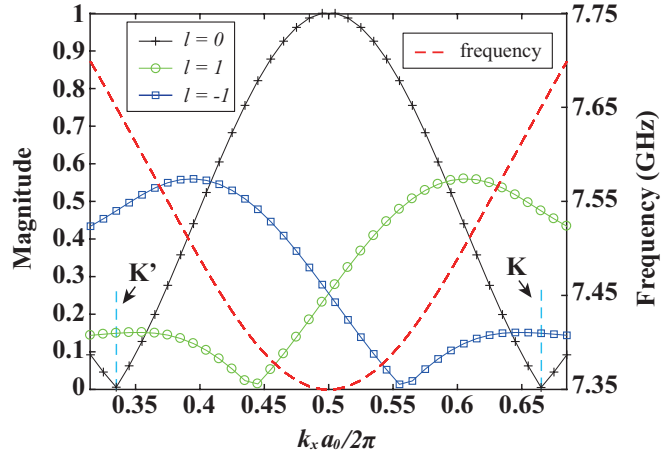}
	\caption{The projection of $E_z$ of the surface states along the white circle in Fig.~\ref{BG_half_E} onto an orthonormal basis, $e^{il\phi}$ (left axis) and their corresponding frequencies (right axis).}
	\label{k_oam}
\end{figure}

To quantify the chirality of the surface states, we extract $E_z$ along a full circular path (the white circle in Fig.~\ref{BG_half_E}) and decompose the extracted complex field into different OAM modes, i.e. onto an orthonormal basis of $e^{il\phi}$ ($l$ is the OAM index). As illustrated in Fig.~\ref{k_oam}, at the $K$ and $K'$ points, the zero-OAM mode is negligible, and for other $k_x$, zero-OAM mode appears. These findings are consistent with the results in Fig.~\ref{BG_half_E}.

The non-zero OAM modes have chirality. Obviously, the left and right branches have opposite chirality. For almost all the states, only one OAM mode dominates. The frequencies of the surface states with respect to $k_x$ can be accessed from the dispersion curve of these states (blue segment in Fig.~\ref{BG_half}) and are also drawn in Fig.~\ref{k_oam}. The advantage of the chiral surface states is that at a specific frequency, the right and left propagating states can be selectively excited by a chiral source. For example, the OAM mode of order $1$ dominates for the surface state at the $K$ point, but is quite weak for the surface state at the $K'$ point. It means that although a source carrying pure OAM of order $1$ can excite both the surface states at the $K$ and $K'$ points, the state excited at the $K$ point will be much stronger. So the rightward propagating wave will be observed and the leftward propagating wave will be very weak. Although there is a zero-OAM mode at most $k_x$, it will have no effect on the sign of excited $k_x$ because it is orthogonal to the source mode. Importantly, the OAM mode is pure at $k_x a_0/ 2\pi =0.445$ and $0.555$, corresponding to $7.39$~GHz. Hence, at $7.39$~GHz, a chiral source carrying an OAM of order $-1$ ($1$) will excite the leftward (rightward) propagating surface state with the best propagation directivity. When a source carries mixed OAM modes that exactly match with the decomposed results at certain $k_x$, the corresponding surface state can also be solely excited. One more thing worth mentioning is that this decomposition scheme does not only apply in the surface states, but can also be used to analyze the propagation directivity of edge states with cladding layers.

\section{Simulation results}

\begin{figure}[htbp]
	\centering
	\includegraphics[width=0.7\columnwidth]{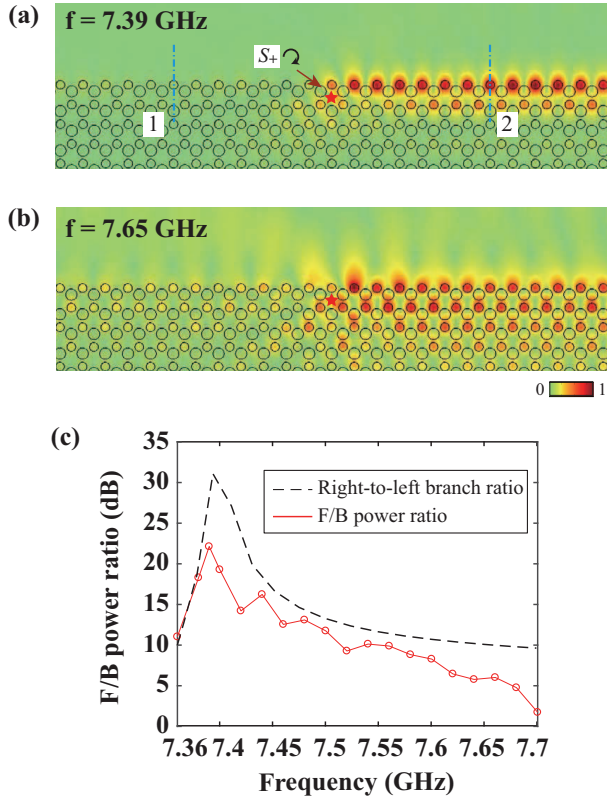}
	\caption{The excited surface states by a six-line-source array carrying OAM of order $1$. The plotting frequencies are (a) $7.39$~GHz and (b) $7.65$~GHz. (c) The comparison between the F/B power ratio and the right-to-left branch ratio.}
	\label{simE}
\end{figure}

Based on our analyses above, the chiral surface states can be excited by a chiral source. In the following, we use a six-line-source array to generate an OAM of order $1$ and put it within the hexagonal cluster as shown in Fig.~\ref{simE}(a). The whole structure is simulated in COMSOL. The simulated amplitudes distributions of $E_z$ at $7.39$~GHz and $7.65$~GHz (the $K$/$K'$ points) are drawn in Figs.~\ref{simE}(a) and (b). The fields being excited have the same profiles as the corresponding eigenstates in Fig.~\ref{BG_half_E}. Meanwhile, as expected, we can observe a nearly perfect unidirectional propagation of the excited surface states at $7.39$~GHz. At $7.65$~GHz, the main power goes rightward, but we can still observe the backward propagation due to the existence of the OAM mode of order $1$ at the $K'$ point. Additionally, because the location of this state on the dispersion diagram is much closer to the light line, the EM field is less localized and extends some distance from the edge sites to air.

Then, we define the propagation directivity by using the forward to backward (F/B) power ratio, $10\text{log}_{10}(U_2/U_1)$. Here, $U_1$ and $U_2$ denote the amount of the EM energy that flows through the lines $1$ and $2$ (indicated by the dashed blue lines in Fig.~\ref{simE}(a)) and it is calculated using $U=1/2 \int_l  \text{Re} (\bf{E} \times \bf{H}^*) \cdot \bf{n} \it{dl}$ ($\bf{n}$ is the normal vector of $1$ and $2$), i.e. by doing the line integral along $1$ and $2$, respectively. Similarly, we define the right-to-left branch ratio as $20\text{log}_{10}(W_{right}/W_{left})$. $W_{right}$ and $W_{left}$ are the magnitudes of  the decomposed OAM modes of order $1$ of the eigenstates on the right and left branches (Fig.~\ref{k_oam}). Not surprisingly, the F/B power ratio has the peak at $7.39$~GHz in Fig.~\ref{simE}(c). The unidirectional propagation is well preserved from $7.36$ to $7.5$~GHz (F/B power ratio larger than $10$~dB). Meanwhile, the The F/B power ratio shows the same trend as the result obtained from the eigenstates, i.e. the right-to-left branch ratio, which validates our local OAM-decomposition method for the analysis of the directivity of surface states. The local peak at $7.44$~GHz from the COMSOL simulations (red solid curve) could possibly be caused by the imperfect absorption of the surface states at the scattering boundaries.

\section{Sharp bends}

\begin{figure}[htbp]
	\centering
	\includegraphics[width=0.7\columnwidth]{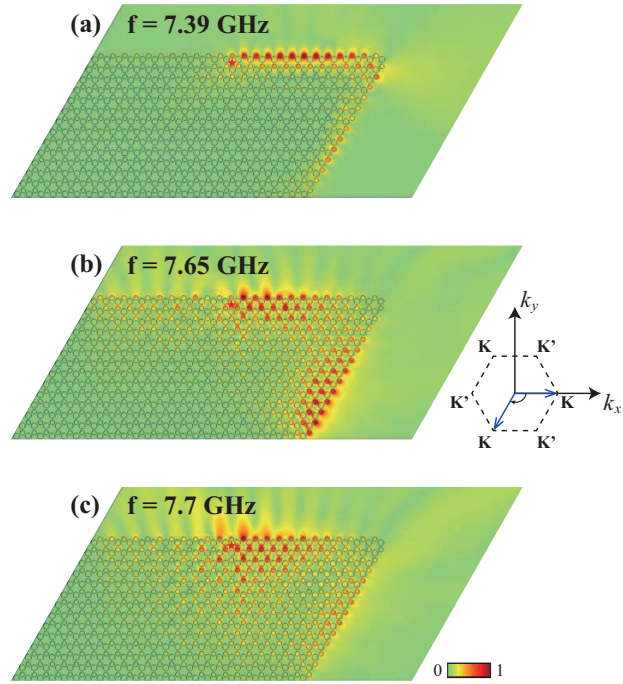}
	\caption{Surface states against sharp bends. The surface sates are excited by a six-line-source array carrying OAM of order $1$. The plotting frequencies are (a) $7.39$~GHz, (b) $7.65$~GHz and (c) $7.7$~GHz.}
	\label{bend}
\end{figure}

We explore the behavior of the chiral surface states when a sharp bend is introduced. In Fig.~\ref{bend}, a $60$-degree bend is made and the same chiral source carrying an OAM of order $1$ is used for excitation. At $7.65$~GHz, i.e. the $K$ point, the wave vectors before and after the corner are drawn in the right panel of Fig.~\ref{bend}(b). The two wave vectors differing by an integer number of reciprocal lattice vectors are considered to be equivalent. Hence, there will be no momentum mismatch before and after the corner and the transmission experiences no loss. This state is near the light line, so the EM fields extend to air, just like the result in Fig.~\ref{simE}(b). Apart from the $K$ point, the symmetry of the crystal cannot compensate for the momentum difference before and after the corner. So, we can see in Figs.~\ref{bend}(a) and (c) that the surface states cannot effectively propagate against the bend. Interestingly, the field distributions at the three frequencies have a common feature that the propagation directivity is maintained when there is a bend. It is obvious for the surface state at $7.39$~GHz: although the power cannot go through the bend, it can neither go back to the source. This finding is consistent with the results in~\cite{Ma2016all}, which results from the topology of the valley PC. The rest of the power is allowed to radiate out to the air at the corner, which can be seen in Figs.~\ref{bend}(a) and (c).

\section{Conclusion}

In conclusion, we have proposed and analyzed the surface states at the zigzag edge of a valley PC. By decomposing the electric field within a hexagonal cluster at the edge into different OAM modes, we identify the chirality of the surface states. By calculating the ratio of the decomposed OAM modes, the leftward and rightward transferred energies are quantified. Furthermore, the decomposition results can help us to selectively excite the left and right propagating surface states. Due to the topology of the valley PC, when a sharp bend is introduced, the chiral surface state exhibits the same propagation directivity as the straight one. The surface states at the $K$ and $K'$ points can propagate through the bend and the others will leak out of the corner.

\section*{Acknowledgments}

Research Grants Council, University Grants Committee (17209918); Asian Office of Aerospace
Research and Development (FA2386-17-1-0010), National Natural Science Foundation of China (61975177, 61271158); HKU Seed Fund (201711159228); Thousand Talents Program for Distinguished Young Scholars of China.

\section*{Disclosures}
The authors declare no conflicts of interest.

\end{document}